\documentclass[twocolumn,superscriptaddress,showpacs,preprintnumbers,amsmath,amssymb]{revtex4}

\usepackage{graphicx,color}
\usepackage{dcolumn}
\usepackage{bm}

\newcommand{\lc}{\left<}
\newcommand{\rc}{\right>}
\newcommand{\lr}{\left|}
\newcommand{\rl}{\right|}
\newcommand{\lb}{\left(}
\newcommand{\rb}{\right)}
\newcommand{\ls}{\left[}
\newcommand{\rs}{\right]}
\newcommand{\Lb}{\left\{}

\newcommand{\ff}[1]{\frac{1}{#1}}

\begin{document}


\title{Isospin corrections for superallowed Fermi $\beta$ decay\\
    in self-consistent relativistic random phase approximation approaches}

\author{Haozhao Liang}
 \affiliation{State Key Lab Nucl. Phys. {\rm\&} Tech., School of Physics, Peking University, Beijing 100871, China}
 \affiliation{Institut de Physique Nucl\'eaire, IN2P3-CNRS and Universit\'e Paris-Sud,
    F-91406 Orsay Cedex, France}

\author{Nguyen Van Giai}
\affiliation{Institut de Physique Nucl\'eaire, IN2P3-CNRS and Universit\'e Paris-Sud,
    F-91406 Orsay Cedex, France}

\author{Jie Meng}
 \affiliation{State Key Lab Nucl. Phys. {\rm\&} Tech., School of Physics, Peking University, Beijing 100871, China}
 \affiliation{Department of Physics, University of Stellenbosch, Stellenbosch, South Africa}

\date{\today}

\begin{abstract}
Self-consistent random phase approximation (RPA) approaches in the
relativistic framework are applied to calculate the isospin
symmetry-breaking corrections $\delta_c$ for the $0^+\rightarrow0^+$
superallowed transitions. It is found that the corrections
$\delta_c$ are sensitive to the proper treatments of the Coulomb
mean field, but not so much to specific effective interactions. With
these corrections $\delta_c$, the nucleus-independent $\mathcal{F}t$
values are obtained in combination with the experimental $ft$ values
in the most recent survey and the improved radiative corrections. It
is found that the constancy of the $\mathcal{F}t$ values is
satisfied for all effective interactions employed. Furthermore, the
element $V_{ud}$ and unitarity of the Cabibbo-Kobayashi-Maskawa
matrix are discussed.
\end{abstract}

\maketitle

\section{Introduction}

The Cabibbo-Kobayashi-Maskawa (CKM) matrix
\cite{Cabibbo1963,Kobayashi1973} relates the quark eigenstates of
the weak interaction with the quark mass eigenstates. The unitarity
condition of the CKM matrix provides a rigorous test for the
standard model description of electroweak interactions. Its leading
matrix element, $V_{ud}$, only depends on the first generation
quarks and so is the element that can be determined most precisely.
There are three traditional methods to determine $|V_{ud}|$
experimentally: nuclear $0^+\rightarrow0^+$ superallowed Fermi
$\beta$ decays \cite{Hardy2005}, neutron decay \cite{Thompson1990},
and pion $\beta$ decay \cite{Pocanic2004}. Recently, experiments
with nuclear mirror transitions provided another independent
sensitive source for extracting the value of $|V_{ud}|$
\cite{Naviliat-Cuncic2009}.

Among these methods, the most precise determination of $|V_{ud}|$
comes from the study of nuclear $0^+\rightarrow0^+$ superallowed
Fermi $\beta$ decays \cite{Amsler2008}. These pure Fermi transitions between nuclear
isobaric analog states (IAS) allow for a direct measurement of the
vector coupling constant $G_V$ of semileptonic weak interactions by
\begin{equation}\label{EQ:GV}
    G_V^2 = \frac{K}{2(1+\Delta^V_R)\mathcal{F}t}.
\end{equation}
Together with the Fermi coupling constant $G_F$ for purely leptonic
decays, the up-down element of the CKM matrix can be determined,
$V_{ud} = G_V/G_F$. In Eq.~(\ref{EQ:GV}), $K/(\hbar
c)^6=2\pi^3\hbar\ln2/(m_ec^2)^5$ and $\Delta^V_R$ is the
transition-independent part of radiative corrections caused, for
example, by the processes where the emitted electron may emit a
bremsstrahlung photon that goes undetected in the experiment
\cite{Marciano2006,Towner2008}. The nucleus-independent
$\mathcal{F}t$ value is obtained by the corrections to the
experimental $ft$ values for radiative effects as well as isospin
symmetry breaking by Coulomb and charge-dependent nuclear forces
\cite{Hardy2005},
\begin{equation}\label{EQ:Ft}
    \mathcal{F}t = ft(1+\delta'_R)(1+\delta_{\rm NS}-\delta_c),
\end{equation}
where $f$ and $t$ represent the statistical rate function and
partial half-life, respectively, and are obtained through
measurements of the $Q$ values, branching ratios, and half-lives for
the superallowed decays. The correction terms $\delta'_R$ and
$\delta_{\rm NS}$ represent the transition-dependent radiative
corrections \cite{Marciano2006,Towner2008}. The correction term
$\delta_c$ is the isospin symmetry-breaking correction, accounting
for the isospin symmetry breaking in nuclei.

The isospin is not an exact symmetry mainly due to the presence of
the Coulomb forces in nuclei. The non-conservation of isospin
symmetry induces a slight reduction of the superallowed transition
strength $|M_F|^2$ from its ideal value $|M_0|^2$:
\begin{equation}\label{EQ:MF}
  |M_F|^2 =
  |\left< f\right|T_\pm\left|i \right>|^2
  = |M_0|^2(1-\delta_c),
\end{equation}
where $M_0 = \sqrt{2}$ for $T = 1$ states with the exact isospin
symmetry.

Shell model calculations are generally used to determine the isospin
symmetry-breaking corrections $\delta_c$. Recently, by including the
core orbitals, an improvement on such corrections has been achieved
and a good agreement among the nucleus-independent $\mathcal{F}t$
values for the 13 well-measured cases has been obtained
\cite{Towner2008}.

Alternatively, the self-consistent random phase approximation (RPA)
based on microscopic mean field theories is another reliable
approach for the superallowed transition strength $M_F$. Such
calculations were performed for a few nuclei with the
non-relativistic Skyrme Hartree-Fock approach in the 1990s
\cite{Sagawa1996}. Since then no further investigation followed even
though significant progress in self-consistent RPA in
charge-exchange channels have been made
\cite{Engel1999,Fracasso2005,DeConti1998,Paar2004,Liang2008}.

During the last decade, great efforts have been dedicated to
developing the charge-exchange (Q)RPA within the relativistic
framework. From the early model which only contains a rather small
configuration space \cite{DeConti1998} to the sophisticated model
which includes Bogoliubov transformation and proton-neutron pairing
\cite{Paar2004}, these approaches are aimed at describing the
spin-isospin resonances, $\beta$ decay rates, neutrino-nucleus cross
sections, etc., in a systematical, reliable, and predictive way.
Recently, based on the success of the newly established
density-dependent relativistic Hartree-Fock (RHF) approach
\cite{Long2006,Long2006b,Sun2008}, a fully self-consistent
charge-exchange RPA has been established and the first applications
have been performed for spin-isospin resonances like Gamow-Teller
and spin-dipole resonances \cite{Liang2008}. A very satisfactory
agreement with the experimental data was obtained without any
readjustment of the energy functional. Therefore, it is appropriate
now to re-investigate the isospin corrections for superallowed Fermi
$\beta$ decay with these relativistic approaches. It is not the aim
here to claim that a covariant framework is necessarily more
appropriate for this problem than a non-relativistic one such as
Skyrme Hartree-Fock plus RPA. The key point, which will be discussed
in Sec. \ref{Sub:deltac}, is a full treatment of Coulomb and nuclear
interactions in both their direct and exchange contributions. In
this respect, satisfactory non-relativistic RPA studies of the
$\delta_c$ corrections are not available.

In this paper, the self-consistent RPA approaches in the
relativistic framework will be applied to calculate the isospin
symmetry-breaking corrections $\delta_c$. With the corrections thus
obtained, the nucleus-independent $\mathcal{F}t$ values will be
obtained in combination with the experimental $ft$ values in the
most recent survey \cite{0812.1202v1} and the improved radiative
corrections \cite{Marciano2006,Towner2008}. The element $V_{ud}$ and
unitarity of the CKM matrix will then be discussed.

\section{Self-consistent Relativistic RPA}

The basic ansatz of the relativistic Hartree (RH), also known as
relativistic mean field (RMF), and relativistic Hartree-Fock (RHF)
theories is a Lagrangian density $\mathcal{L}$, where nucleons are
described as Dirac spinors that interact via the exchange of
$\sigma$-, $\omega$-, $\rho$-, $\pi$-mesons and the photon
\cite{Serot1986,Bouyssy1987,Meng2006}. In order to give a
satisfactory description of nuclear matter and finite nuclei, the
nonlinear self-coupling of mesons, e.g., in
Refs.~\cite{Lalazissis1997,Sugahara1994,Long2004}, or
density-dependent meson-nucleon couplings, e.g., in
Refs.~\cite{Niksic2002,Long2006}, are introduced.

The effective Hamiltonian operator $\hat H$ can be obtained with the
general Legendre transformation. Together with the trial ground
state (Slater determinant) in the Hartree or Hartree-Fock
approximation, the energy functional can be written as
\begin{eqnarray}\label{EF}
    E &=& \lc \Phi_0 \lr \hat H \rl \Phi_0 \rc\nonumber\\
    &=& \sum_a \lc a \rl{\mathbf{\alpha}\cdot\mathbf{p}+\beta M}\lr a \rc
        +\ff2\sum_{ab}\lc{ab}\rl{V(1,2)}\lr{ba}\rc\nonumber\\
        &&-\ff2\sum_{ab}\lc{ab}\rl{V(1,2)}\lr{ab}\rc,
\end{eqnarray}
where the first term is the kinetic energy, the second and the last
terms are the direct (Hartree) and exchange (Fock) energies,
respectively. In the Hartree approximation, the Fock term is
neglected for simplicity. The two-body interaction $V(1,2)$ includes
the following meson-nucleon and photon-nucleon interactions:
\begin{subequations}
\begin{eqnarray}
    V_\sigma(1,2) &=& -[g_\sigma\gamma_0]_1[g_\sigma\gamma_0]_2D_\sigma(1,2),\label{sigma}\\
    V_\omega(1,2) &=& [g_\omega\gamma_0\gamma^\mu]_1[g_\omega\gamma_0\gamma_\mu]_2D_\omega(1,2),\label{omega}\\
    V_\rho(1,2) &=& [g_\rho\gamma_0\gamma^\mu\vec\tau]_1\cdot [g_\rho\gamma_0\gamma_\mu\vec\tau]_2D_\rho(1,2),\label{rho}\\
    V_\pi(1,2) &=& -\ls\frac{f_\pi}{m_\pi}\vec\tau\gamma_0\gamma_5\gamma^k\partial_k\rs_1\cdot\nonumber\\
         &&\ls\frac{f_\pi}{m_\pi}\vec\tau\gamma_0\gamma_5\gamma^l\partial_l\rs_2
         D_\pi(1,2),\label{pion}\\
    V_A(1,2) &=& \frac{e^2}{4}[\gamma_0\gamma^\mu(1-\tau_3)]_1[\gamma_0\gamma_\mu(1-\tau_3)]_2D_A(1,2),\label{photon}\nonumber\\
\end{eqnarray}
\end{subequations}
with the finite-range Yukawa type propagator
\begin{equation}\label{Yukawa}
    D_i(1,2) = \frac{1}{4\pi}\frac{e^{-m_i|\mathbf{r}_1-\mathbf{r}_2|}}
        {|\mathbf{r}_1-\mathbf{r}_2|}.
\end{equation}
Furthermore, in order to cancel the contact interaction coming from
the pion pseudovector coupling, a zero-range pionic counterterm
should be included \cite{Bouyssy1987,Liang2008}:
\begin{equation}\label{counter}
    V_{\pi\delta}(1,2)=g'\ls\frac{f_\pi}{m_\pi}\vec\tau\gamma_0\gamma_5\boldsymbol\gamma\rs_1\cdot
         \ls\frac{f_\pi}{m_\pi}\vec\tau\gamma_0\gamma_5\boldsymbol\gamma\rs_2
         \delta(\mathbf{r}_1-\mathbf{r}_2),
\end{equation}
with $g'=1/3$. Thus, $g'$ is not an adjustable parameter.

The RPA equations can be obtained by taking the second derivative of
the energy functional $E$. In the charge-exchange channels, the RPA
equations become
\begin{equation}\label{RPA}
    \lb\begin{array}{cc}
        \mathcal A^J_{p\bar np'\bar n'} & \mathcal B^J_{p\bar nn'\bar p'} \\
        -\mathcal B^J_{n\bar pp'\bar n'} & -\mathcal A^J_{n\bar pn'\bar p'}
    \end{array}\rb
    \lb\begin{array}{c} U^{J\nu}_{p'\bar n'} \\ V^{J\nu}_{n'\bar p'} \end{array}\rb
    =\omega_\nu
    \lb\begin{array}{c} U^{J\nu}_{p\bar n} \\ V^{J\nu}_{n\bar p} \end{array}\rb,
\end{equation}
where $p$ and $\bar p$ ($n$ and $\bar n$) denote unoccupied and
occupied proton (neutron) states. These equations describe both the
$T_+$ and $T_-$ channels. It should be emphasized that the
unoccupied states include not only the states above the Fermi
surface, but also the states in the Dirac sea. The RPA matrices
$\mathcal A$ and $\mathcal B$ read
\begin{subequations}\label{AB}
\begin{align}
    \mathcal{A}_{12,34} &= (E_1-E_2)\delta_{12,34}+\lc{14}\rl{V_{\rm ph}}\lr{32-23}\rc,\\
    \mathcal{B}_{12,34} &= -\lc{13}\rl{V_{\rm ph}}\lr{42-24}\rc,
\end{align}
\end{subequations}
where the first term in the ket represents the direct contribution,
and the second term represents the exchange contribution. In the RPA
built on the Hartree mean field, the exchange contributions in
Eqs.~(\ref{AB}) are accordingly neglected.

In the self-consistent RPA calculations, the particle-hole residual
interaction $V_{\rm ph}$ should be derived from the same energy
functional $E$ as that used in the ground-state description. The
explicit density dependence of the meson-nucleon couplings
introduces, in principle, additional rearrangement terms in the
particle-hole residual interaction $V_{\rm ph}$, and their
contributions are essential for a quantitative description of
excited states \cite{Niksic2002a}. However, since the rearrangement
terms are due to the dependence on isoscalar ground-state densities,
it is easy to see that they are absent in the charge-exchange
channels. Therefore, in the description of superallowed Fermi
$\beta$ decays, the particle-hole residual interaction $V_{\rm ph}$
is just the meson-nucleon interactions shown in
Eqs.~(\ref{sigma})-(\ref{pion}) and (\ref{counter}). The
photon-nucleon interaction in Eq.~(\ref{photon}) does not contribute
to the particle-hole residual interaction because the configurations
are of the neutron-proton type.

The eigenvectors of the RPA equations (\ref{RPA}) are separated into
two groups, which respectively represent the excitations of the
$T_-$ and $T_+$ channels with the following normalization
conditions:
\begin{equation}
    \Lb\begin{array}{ll}
        \sum_{p\bar n}(U^{J\nu}_{p\bar n})^2-\sum_{n\bar p}(V^{J\nu}_{n\bar p})^2=+1, &
        \mbox{for $T_-$ channel,}\\
        \sum_{p\bar n}(U^{J\nu}_{p\bar n})^2-\sum_{n\bar p}(V^{J\nu}_{n\bar p})^2=-1, &
        \mbox{for $T_+$ channel.}
    \end{array}\right.
\end{equation}
Then, the excitation energies and $X$, $Y$ amplitudes in the $T_-$
channel read
\begin{equation}
    \Omega_\nu=+\omega_\nu,\quad X^{J\nu}_{p\bar n}=U^{J\nu}_{p\bar n},\quad Y^{J\nu}_{n\bar p}=V^{J\nu}_{n\bar p},
\end{equation}
whereas the excitation energies and $X$, $Y$ amplitudes in the $T_+$
channel are
\begin{equation}
    \Omega_\nu=-\omega_\nu,\quad X^{J\nu}_{n\bar p}=V^{J\nu}_{n\bar p},\quad Y^{J\nu}_{p\bar n}=U^{J\nu}_{p\bar n}.
\end{equation}

The $0^+\rightarrow0^+$ superallowed transition operators are $T_-$
or $T_+$. The transition probabilities between the ground-state and
excited states read
\begin{subequations}
\begin{align}
    B^-_{J\nu} &=
        \left|\sum_{p\bar n}X^{J\nu}_{p\bar n}\lc p\right.||{T_-}||\left.{\bar
        n}\rc+\sum_{n\bar p}(-)^{j_n+j_{\bar p}}Y^{J\nu}_{n\bar p}
        \lc {\bar p}\right.||{T_-}||\left.{n}\rc\right|^2,\\
    B^+_{J\nu} &=
        \left|\sum_{n\bar p}X^{J\nu}_{n\bar p}\lc n\right.||{T_+}||\left.{\bar p}\rc
        +\sum_{p\bar n}(-)^{j_p+j_{\bar n}}Y^{J\nu}_{p\bar n}
        \lc{\bar n}\right.||{T_+}||\left.{p}\rc\right|^2.
\end{align}
\end{subequations}

Before ending this section, it is worthwhile to make the following
remark about the self-consistency of the RH+RPA approach when it is
applied to the $0^+\rightarrow0^+$ transitions. Within this
approach, it is known that, in order to reproduce the excitation
energies of Gamow-Teller resonances, one has to adjust the $\pi NN$
particle-hole residual interaction and that $g'$ cannot be kept
equal to $1/3$ \cite{DeConti1998,Paar2004}. However, for the
$0^+\rightarrow0^+$ channel in the present paper, the direct
contributions from the pion vanish. Therefore, in this sense, the
self-consistency is also fulfilled in RH+RPA approach for the
superallowed Fermi $\beta$ decays.

\section{Results and Discussion}

For all the calculations in this paper, the spherical symmetry is
assumed and the filling approximation is applied to the last
partially occupied orbital. The radial Dirac equations are solved in
coordinate space by the Runge-Kutta method within a spherical box
with a box radius $R=15$~fm and a mesh size $dr=0.1$~fm
\cite{Meng1998}. The single-particle wave functions thus obtained
are used to construct the RPA matrices $\mathcal A$ and $\mathcal B$
in Eqs.~(\ref{AB}) with the single-particle energy truncation $[-M,
M+120~{\rm MeV}]$, i.e., the occupied states are the positive energy
states below the Fermi surface, whereas the unoccupied states can be
either positive energy states above the Fermi surface or bound
negative energy states \cite{Liang2008}. With these numerical
inputs, the model-independent sum rule,
\begin{equation}\label{sumrule}
    \sum_\nu B^-_\nu - \sum_\nu B^+_\nu = N - Z,
\end{equation}
can be fulfilled up to $10^{-5}$ accuracy, and the isospin
symmetry-breaking corrections $\delta_c$ are stable with respect to
these numerical inputs at the same level of accuracy.

\subsection{Isospin symmetry-breaking correction $\delta_c$}\label{Sub:deltac}

\begin{table*}
\caption{Isospin symmetry-breaking corrections $\delta_c$
    for the $0^+\rightarrow0^+$ superallowed transitions
    obtained by self-consistent RHF+RPA calculations
    with PKO1 \cite{Long2006}, PKO2 \cite{Long2008}, and PKO3
    \cite{Long2008}
    as well as self-consistent RH+RPA calculations
    with DD-ME1 \cite{Niksic2002}, DD-ME2 \cite{Lalazissis2005},
    NL3 \cite{Lalazissis1997}, and TM1 \cite{Sugahara1994}.
    The column PKO1* presents the results obtained with PKO1
    without the Coulomb exchange (Fock) term.
    The results obtained by shell model calculations \cite{Towner2008}
    are listed in the column T\&H for comparison.
    All values are expressed in percents.
    \label{Tab1}}
\begin{ruledtabular}
    \begin{tabular}{lccccccccl}
    & PKO1 & PKO2 & PKO3 & PKO1* & DD-ME1 & DD-ME2 & NL3 & TM1 & T\&H \cite{Towner2008} \\ \hline
        $^{10}$C  $\rightarrow$ $^{10}$B  & 0.082 & 0.083 & 0.088 & 0.148 & 0.149 & 0.150 & 0.124 & 0.133 & 0.175(18) \\
        $^{14}$O  $\rightarrow$ $^{14}$N  & 0.114 & 0.134 & 0.110 & 0.178 & 0.189 & 0.197 & 0.181 & 0.159 & 0.330(25) \\
        $^{18}$Ne $\rightarrow$ $^{18}$F  & 0.270 & 0.277 & 0.288 & 0.357 & 0.424 & 0.430 & 0.344 & 0.373 & 0.565(39) \\
        $^{26}$Si $\rightarrow$ $^{26}$Al & 0.176 & 0.176 & 0.184 & 0.246 & 0.252 & 0.252 & 0.213 & 0.226 & 0.435(27) \\
        $^{30}$S  $\rightarrow$ $^{30}$P  & 0.497 & 0.550 & 0.507 & 0.625 & 0.612 & 0.633 & 0.551 & 0.648 & 0.855(28) \\
        $^{34}$Ar $\rightarrow$ $^{34}$Cl & 0.268 & 0.281 & 0.267 & 0.359 & 0.368 & 0.376 & 0.438 & 0.320 & 0.665(56) \\
        $^{38}$Ca $\rightarrow$ $^{38}$K  & 0.313 & 0.330 & 0.313 & 0.406 & 0.431 & 0.441 & 0.390 & 0.572 & 0.765(71) \\
        $^{42}$Ti $\rightarrow$ $^{42}$Sc & 0.384 & 0.387 & 0.390 & 0.460 & 0.515 & 0.523 & 0.436 & 0.443 & 0.935(78) \\
        $^{26}$Al $\rightarrow$ $^{26}$Mg & 0.139 & 0.138 & 0.144 & 0.193 & 0.198 & 0.198 & 0.172 & 0.179 & 0.310(18) \\
        $^{34}$Cl $\rightarrow$ $^{34}$S  & 0.234 & 0.242 & 0.231 & 0.298 & 0.302 & 0.307 & 0.289 & 0.267 & 0.650(46) \\
        $^{38}$K  $\rightarrow$ $^{38}$Ar & 0.278 & 0.290 & 0.276 & 0.344 & 0.363 & 0.371 & 0.334 & 0.484 & 0.655(59) \\
        $^{42}$Sc $\rightarrow$ $^{42}$Ca & 0.333 & 0.334 & 0.336 & 0.395 & 0.442 & 0.448 & 0.377 & 0.383 & 0.665(56) \\
        $^{54}$Co $\rightarrow$ $^{54}$Fe & 0.319 & 0.317 & 0.321 & 0.392 & 0.395 & 0.393 & 0.355 & 0.368 & 0.770(67) \\
        $^{66}$As $\rightarrow$ $^{66}$Ge & 0.475 & 0.475 & 0.469 & 0.571 & 0.568 & 0.572 & 0.560 & 0.524 & 1.56(40)  \\
        $^{70}$Br $\rightarrow$ $^{70}$Se & 1.140 & 1.118 & 1.107 & 1.234 & 1.232 & 1.268 & 1.230 & 1.226 & 1.60(25)  \\
        $^{74}$Rb $\rightarrow$ $^{74}$Kr & 1.088 & 1.091 & 1.071 & 1.230 & 1.233 & 1.258 & 1.191 & 1.234 & 1.63(31)  \\
    \end{tabular}
\end{ruledtabular}
\end{table*}

\begin{table}
\caption{Excitation energies $E_x$ for the $0^+\rightarrow0^+$
    superallowed transitions
    measured by taking the ground state of the corresponding even-even nuclei as
    reference. In the comparison with the experimental values taken from
    the recent survey \cite{0812.1202v1}, the corrections due to the
    proton-neutron mass difference in particle-hole configurations are
    made for the calculated results.
    All units are in MeV.
    \label{Tab2}}
\begin{ruledtabular}
    \begin{tabular}{lcccc}
    & expt. & PKO1 & PKO1* & DD-ME2 \\ \hline
        $^{10}$C  $\rightarrow$ $^{10}$B  & -1.908 & -1.698 & -2.307 & -2.236 \\
        $^{14}$O  $\rightarrow$ $^{14}$N  & -2.831 & -2.420 & -2.989 & -3.081 \\
        $^{18}$Ne $\rightarrow$ $^{18}$F  & -3.402 & -3.195 & -3.497 & -3.451 \\
        $^{26}$Si $\rightarrow$ $^{26}$Al & -4.842 & -4.531 & -5.139 & -5.110 \\
        $^{30}$S  $\rightarrow$ $^{30}$P  & -5.460 & -4.845 & -5.326 & -5.395 \\
        $^{34}$Ar $\rightarrow$ $^{34}$Cl & -6.063 & -5.559 & -6.129 & -6.278 \\
        $^{38}$Ca $\rightarrow$ $^{38}$K  & -6.612 & -6.035 & -6.611 & -6.775 \\
        $^{42}$Ti $\rightarrow$ $^{42}$Sc & -7.000 & -6.661 & -6.970 & -6.964 \\
        $^{26}$Al $\rightarrow$ $^{26}$Mg & ~4.233 & ~3.908 & ~4.372 & ~4.350 \\
        $^{34}$Cl $\rightarrow$ $^{34}$S  & ~5.492 & ~5.062 & ~5.428 & ~5.561 \\
        $^{38}$K  $\rightarrow$ $^{38}$Ar & ~6.044 & ~5.557 & ~5.936 & ~6.083 \\
        $^{42}$Sc $\rightarrow$ $^{42}$Ca & ~6.426 & ~6.118 & ~6.333 & ~6.333 \\
        $^{54}$Co $\rightarrow$ $^{54}$Fe & ~8.244 & ~7.720 & ~8.221 & ~8.240 \\
        $^{66}$As $\rightarrow$ $^{66}$Ge & ~9.579 & ~9.044 & ~9.488 & ~9.677 \\
        $^{70}$Br $\rightarrow$ $^{70}$Se & ~9.970 & ~9.632 & ~9.805 & ~9.852 \\
        $^{74}$Rb $\rightarrow$ $^{74}$Kr & 10.417 & 10.005 & 10.349 & 10.437 \\
    \end{tabular}
\end{ruledtabular}
\end{table}

In Table~\ref{Tab1}, the isospin symmetry-breaking corrections
$\delta_c$ in Eq.~(\ref{EQ:MF}) for the $0^+\rightarrow0^+$
superallowed transitions are shown. The results are obtained by
self-consistent RHF+RPA calculations with PKO1 \cite{Long2006}, PKO2
\cite{Long2008}, PKO3 \cite{Long2008} effective interactions, as
well as by self-consistent RH+RPA calculations with DD-ME1
\cite{Niksic2002}, DD-ME2 \cite{Lalazissis2005}, NL3
\cite{Lalazissis1997}, TM1 \cite{Sugahara1994} effective
interactions. The results obtained by shell model calculations
(T\&H) \cite{Towner2008} are also listed for comparison. The present
corrections $\delta_c$ range from about $0.1\%$ for the lightest
nucleus $^{10}$C to about $1.2\%$ for the heaviest nucleus
$^{74}$Rb, which are 2-3 times smaller than the T\&H results. It is
noticed that even smaller values of $\delta_c$ compared to the shell
model calculations have been recently obtained in Ref.
\cite{Auerbach2009} using perturbation theory. In addition, in
Table~\ref{Tab2} the excitation energies $E_x$ for the
$0^+\rightarrow0^+$ superallowed transitions corresponding to PKO1
and DD-ME2 are shown as examples. These energies are measured by
taking the ground state of the corresponding even-even nuclei as
reference. In the comparison with the experimental values taken from
the recent survey \cite{0812.1202v1}, the corrections due to the
proton-neutron mass difference in particle-hole configurations are
made for the calculated results. A good agreement between the data
and the calculated ones can be seen in Table~\ref{Tab2}.

In Table~\ref{Tab1}, it is found that the present isospin
symmetry-breaking corrections $\delta_c$ for each nucleus can be
unambiguously divided into two categories, those obtained by RHF+RPA
calculations and those obtained by RH+RPA calculations. Comparing
these two categories, it is seen that the corrections $\delta_c$ of
RHF+RPA are systematically smaller than those of RH+RPA. On the
other hand, it is also found that within one category the
corrections $\delta_c$ are not sensitive to specific effective
interactions or the structure of the Lagrangian density. For
instance, within the RH+RPA framework, both the Lagrangian densities
with density-dependent meson-nucleon couplings (DD-ME1, DD-ME2) or
with non-linear meson couplings (NL3, TM1) lead to quite similar
results.

To understand this systematic discrepancy between RHF+RPA and
RH+RPA, it must be kept in mind that in RHF+RPA the exchange (Fock)
terms of mesons and photon are kept in both the mean field and RPA
levels, whereas they are neglected altogether in RH+RPA. Among all
the Fock terms, we expect, in particular, the exchange terms of the
Coulomb field to play an important role due to the following reason.
The IAS would be degenerate with its isobaric multiplet partners,
i.e., $E_x=0$, and it would contain 100\% of the model-independent
sum rule (\ref{sumrule}), i.e., $\delta_c=0$, if the nuclear
Hamiltonian commutes with the isospin raising and lowering operators
$T_\pm$. This would be true when the Coulomb field is switched off.
While this degeneracy is broken by the mean field approximation, no
matter the exchange terms of mesons are included or not, it can be
restored by the RPA calculations as long as the RPA calculations are
self-consistent \cite{Engelbrecht1970}. Therefore, the Coulomb field
is essential for the $0^+\rightarrow0^+$ superallowed transitions
and the Coulomb exchange (Fock) term should be responsible for the
the different isospin symmetry-breaking corrections $\delta_c$ in
RHF+RPA and RH+RPA approaches.

In order to verify the above argument, we have performed the
following calculations. Using PKO1, the Hartree-Fock calculations
are performed by switching off the exchange contributions of the
Coulomb field. From the single-particle spectra thus obtained,
self-consistent RPA calculations are then performed. One may notice
that in such calculations some nuclear properties including binding
energies and rms radii can no longer be reproduced. However, this
does not hinder us from discussing the physics we are concerned
with. The isospin symmetry-breaking corrections $\delta_c$ and the
excitation energies $E_x$ thus obtained are listed in the column
denoted as PKO1* in Tables~\ref{Tab1} and \ref{Tab2}. It is seen
that these results are almost the same as those of RH+RPA
calculations with DD-ME1, DD-ME2, NL3, and TM1, i.e., by switching
off the exchange contributions of the Coulomb field, $E_x$ and
$\delta_c$ in RHF+RPA calculations recover the results in RH+RPA
calculations. In other words, although the meson exchange terms can
be somehow effectively included by adjusting the parameters in the
direct terms, this has not been done for the Coulomb part in the
usual RH approximation.

Therefore, one can conclude that the proper treatments of the
Coulomb field is very important to extract the isospin
symmetry-breaking corrections $\delta_c$.

\subsection{Nucleus-independent $\mathcal{F}t$ values}

\begin{table*}
\caption{Nucleus-independent $\mathcal{F}t$ values.
    The average $\overline{\mathcal{F}t}$ value and the normalized $\chi^2/\nu$
    appear at the bottom. All units are in s.
    \label{Tab3}}
\begin{ruledtabular}
    \begin{tabular}{lcccccccc}
        & PKO1 & PKO2 & PKO3 & PKO1* & DD-ME1 & DD-ME2 & NL3 & TM1 \\ \hline
        $^{10}$C  $\rightarrow$ $^{10}$B  & 3079.6(45) & 3079.5(45) & 3079.4(45) & 3077.5(45) & 3077.5(45) & 3077.5(45) & 3078.3(45) & 3078.0(45) \\
        $^{14}$O  $\rightarrow$ $^{14}$N  & 3078.2(31) & 3077.5(31) & 3078.3(31) & 3076.2(31) & 3075.8(31) & 3075.6(31) & 3076.1(31) & 3076.8(31) \\
        $^{34}$Ar $\rightarrow$ $^{34}$Cl & 3081.9(84) & 3081.5(84) & 3082.0(84) & 3079.1(84) & 3078.8(84) & 3078.6(84) & 3076.7(83) & 3080.3(84) \\
        $^{26}$Al $\rightarrow$ $^{26}$Mg & 3077.7(13) & 3077.7(13) & 3077.5(13) & 3076.0(13) & 3075.8(13) & 3075.8(13) & 3076.6(13) & 3076.4(13) \\
        $^{34}$Cl $\rightarrow$ $^{34}$S  & 3083.5(16) & 3083.3(16) & 3083.6(16) & 3081.6(16) & 3081.4(16) & 3081.3(16) & 3081.8(16) & 3082.5(16) \\
        $^{38}$K  $\rightarrow$ $^{38}$Ar & 3084.1(16) & 3083.8(16) & 3084.2(16) & 3082.1(16) & 3081.5(16) & 3081.3(16) & 3082.4(16) & 3077.8(16) \\
        $^{42}$Sc $\rightarrow$ $^{42}$Ca & 3082.7(21) & 3082.6(21) & 3082.6(21) & 3080.7(21) & 3079.3(21) & 3079.1(21) & 3081.3(21) & 3081.1(21) \\
        $^{54}$Co $\rightarrow$ $^{54}$Fe & 3083.9(27) & 3083.9(27) & 3083.8(27) & 3081.6(27) & 3081.5(27) & 3081.6(27) & 3082.7(27) & 3082.4(27) \\
        $^{74}$Rb $\rightarrow$ $^{74}$Kr & 3094.8(87) & 3094.7(87) & 3095.3(87) & 3090.3(87) & 3090.2(87) & 3089.4(87) & 3091.5(87) & 3090.2(87) \\ \hline
        average                           & 3081.4(7)~ & 3081.3(7)~ & 3081.4(7)~ & 3079.5(7)~ & 3079.1(7)~ & 3079.0(7)~ & 3080.0(7)~ & 3079.1(7)~ \\
        $\chi^2/\nu$                      & ~~~1.1~~~~ & ~~~1.1~~~~ & ~~~1.1~~~~ & ~~~1.0~~~~ & ~~~1.0~~~~ & ~~~1.0~~~~ & ~~~1.0~~~~ & ~~~1.0~~~~ \\
    \end{tabular}
\end{ruledtabular}
\end{table*}

\begin{figure*}
\includegraphics[width=0.6\textwidth]{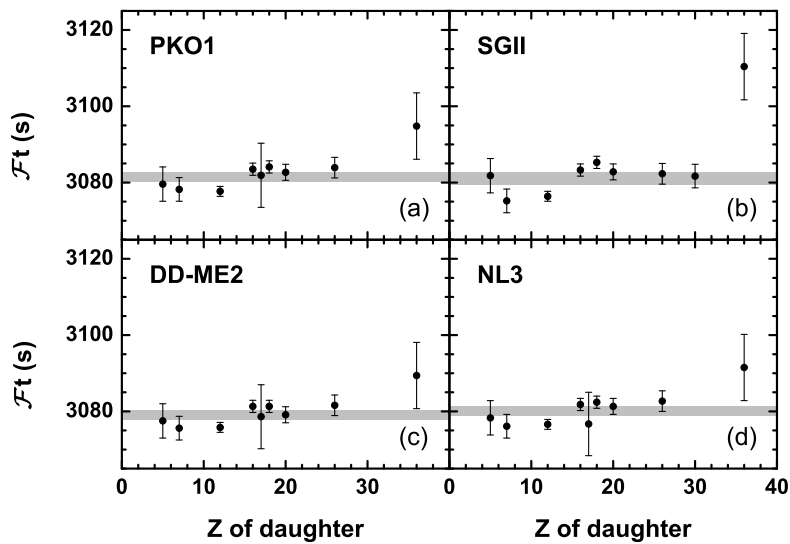}
\caption{Nucleus-independent $\mathcal{F}t$ values as a function of
    the charge $Z$ for the daughter nucleus. The values of $\delta_c$ are respectively obtained
    by RHF+RPA calculations with PKO1 (a),
    by RH+RPA calculations with DD-ME2 (c) and
    NL3 (d),
    as well as by SHF+RPA calculations with SGII \cite{Sagawa1996} (b).
    The shaded horizontal band gives one standard deviation around the average
    $\overline{\mathcal{F}t}$ value.
    \label{Fig1}}
\end{figure*}

Among the $0^+\rightarrow0^+$ superallowed transitions listed in
Table~\ref{Tab1}, some of their measured $ft$ values are summarized
in a recent survey \cite{0812.1202v1}. To obtain the
nucleus-independent $\mathcal{F}t$ values from each experimental
$ft$ value, apart from the isospin symmetry-breaking corrections
$\delta_c$ in Table~\ref{Tab1}, one still needs the values of the
transition-dependent radiative corrections $\delta'_R$ and
nuclear-structure-dependent radiative corrections $\delta_{\rm NS}$.

Using the $\delta'_R$ and $\delta_{\rm NS}$ values from recent
calculations \cite{Towner2008}, $\delta_c$ in Table~\ref{Tab1}, and
measured $ft$ values \cite{0812.1202v1}, the nucleus-independent
$\mathcal{F}t$ values for superallowed Fermi $\beta$ decays are
listed in Table~\ref{Tab3} together with the average
$\overline{\mathcal{F}t}$ values and the values of chi-square per
degree of freedom $\chi^2/\nu$, in which the uncertainty of
$\delta_c$ is taken as zero.

It is found that the chi-square per degree of freedom $\chi^2/\nu$
is $1.0\sim1.1$~s for all effective interactions employed. This
indicates the constancy of the nucleus-independent $\mathcal{F}t$
values is satisfied, even though not as well as the shell model
calculations in Ref.~\cite{0812.1202v1}. It is also found that the
$\mathcal{F}t$ values of RHF+RPA are about 2~s larger than those of
RH+RPA, which are larger than the difference due to the different
effective interactions in either RHF or RH approximations.

The results of RHF+RPA with PKO1, RH+RPA with DD-ME2 and NL3 are
plotted as a function of the charge $Z$ for the daughter nucleus in
Fig.~\ref{Fig1} to illustrate the constancy of the
nucleus-independent $\mathcal{F}t$ values. The shaded horizontal
band gives the standard deviation, which combines the statistical
errors and $\chi^2/\nu$, around the average
$\overline{\mathcal{F}t}$ value.

In order to get a deeper understanding on the treatment of the
Coulomb field, the $\mathcal{F}t$ values from RPA calculations using
Skyrme Hartree-Fock (SHF) with SGII effective interaction are shown
in panel (b) of Fig.~\ref{Fig1}, in which the isospin
symmetry-breaking corrections $\delta_c$ are taken from the Table~I
in Ref.~\cite{Sagawa1996}. It should be emphasized that in these
results the exchange contributions to the Coulomb mean field are
treated in Slater approximation. Although this model leads to a
similar average $\mathcal{F}t$ value,
$\overline{\mathcal{F}t}=3081.1(7)$~s, it is found that the
chi-square per degree of freedom $\chi^2/\nu=1.5$, i.e., the
constancy of the $\mathcal{F}t$ values in this SHF framework is not
as good as that given by the relativistic calculations. In
particular, the $\mathcal{F}t$ value deduced from the nucleus
$^{74}$Rb is seriously overestimated.

\subsection{CKM matrix}

With the nucleus-independent $\mathcal{F}t$ value, the element
$V_{ud}$ of the CKM matrix can be calculated by
\begin{equation}
    V^2_{ud} = \frac{K}{2G^2_F(1+\Delta^V_R)\overline{\mathcal{F}t}},
\end{equation}
where $K/(\hbar c)^6=8120.2787(11)\times 10^{-10}~{\rm GeV}^{-4}{\rm
s}$, $G_F/(\hbar c)^3=1.16637(1)\times10^{-5}~{\rm GeV}^{-2}$
\cite{Amsler2008}, and $\Delta^V_R=2.361(38)\%$ \cite{Towner2008}.
Then in combination with the other two CKM matrix elements
$|V_{us}|=0.2255(19)$ and $|V_{ub}|=0.00393(36)$ \cite{Amsler2008},
one can test the unitarity of the matrix.

\begin{table}
\caption{The element $V_{ud}$ and the sum of squared top-row
elements of the CKM matrix.
    \label{Tab4}}
\begin{ruledtabular}
    \begin{tabular}{lcc}
        & $|V_{ud}|$ & $|V_{ud}|^2+|V_{us}|^2+|V_{ub}|^2$ \\ \hline
        PKO1   & 0.97273(27) & 0.9971(10) \\
        PKO2   & 0.97275(27) & 0.9971(10) \\
        PKO3   & 0.97273(27) & 0.9971(10) \\
        PKO1*  & 0.97303(26) & 0.9977(10) \\
        DD-ME1 & 0.97309(26) & 0.9978(10) \\
        DD-ME2 & 0.97311(26) & 0.9978(10) \\
        NL3    & 0.97295(26) & 0.9975(10) \\
        TM1    & 0.97309(26) & 0.9978(10) \\
    \end{tabular}
\end{ruledtabular}
\end{table}

\begin{figure}
\includegraphics[width=0.45\textwidth]{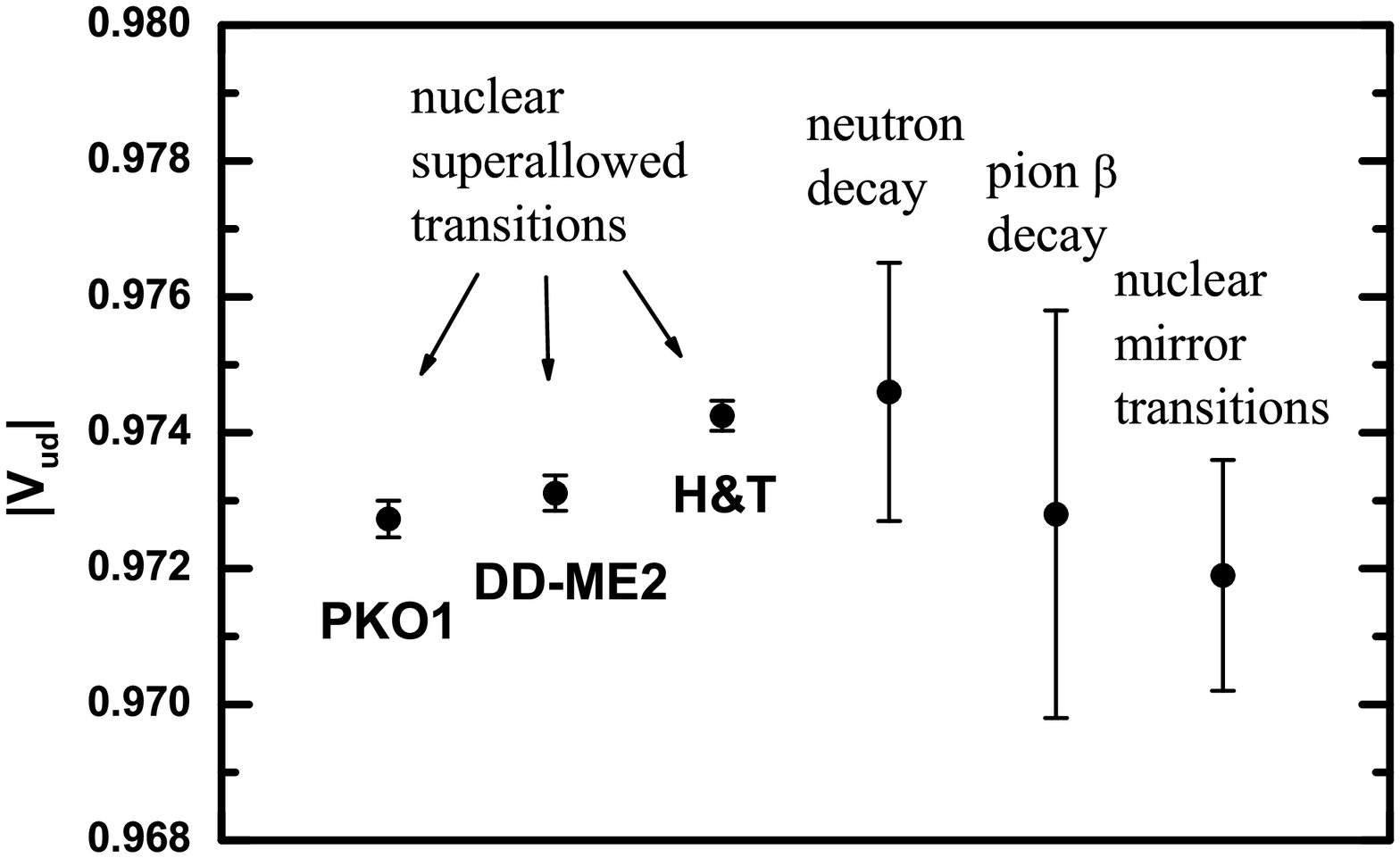}
\caption{The element $V_{ud}$ of the CKM matrix
    obtained by RHF+RPA calculations with PKO1
    and by RH+RPA calculations with DD-ME2 in comparison with
    those in shell model (H\&T) \cite{0812.1202v1} as well as
    in neutron decay \cite{Amsler2008},
    pion $\beta$ decay \cite{Pocanic2004} and nuclear mirror transitions
    \cite{Naviliat-Cuncic2009}.
    \label{Fig2}}
\end{figure}

The element $V_{ud}$  as well as the sum of squared top-row elements
of the CKM matrix are listed in Table~\ref{Tab4}. The uncertainties
of the present results are underestimated to some extent as the
uncertainty of $\delta_c$ is assumed to be zero and the systematic
errors are not taken into account. In Fig.~\ref{Fig2}, the element
$V_{ud}$ of the CKM matrix obtained by RHF+RPA calculations with
PKO1 and by RH+RPA calculations with DD-ME2 are shown in comparison
with those in the shell model (H\&T) \cite{0812.1202v1} as well as
in neutron decay \cite{Amsler2008}, pion $\beta$ decay
\cite{Pocanic2004}, and nuclear mirror transitions
\cite{Naviliat-Cuncic2009}.

It can be clearly seen in Table~\ref{Tab4} that the matrix element
$|V_{ud}|$ determined by the $0^+\rightarrow0^+$ superallowed
transitions mainly depends on the treatment of the Coulomb field and
less sensitive to the particular effective interactions. Switching
either on or off the exchange contributions of the Coulomb field,
the discrepancy caused by different effective interactions is much
smaller than the statistic deviation. It is interesting to note that
the present $|V_{ud}|$ values well agree with those obtained in
neutron decay, pion $\beta$ decay and nuclear mirror transitions.
However, the sum of squared top-row elements considerably deviates
from the unitarity condition, which is in contradiction with the
conclusion in shell model calculations (H\&T) \cite{0812.1202v1}.
This calls for more intensive investigations in the future. For
example, mean field and RPA calculations including the proper
neutron-proton mass difference, isoscalar and isovector pairing, and
deformation should be done. It should also be emphasized that apart
from the proper treatment of pairing by either BCS or Bogoliubov
approaches, the particle number projection must be implemented as
well in order to remove the artificial isospin symmetry breaking
effects due to the particle number violation.

\section{Summary and Perspectives}

In summary, self-consistent relativistic RPA approaches are applied
to calculate the isospin symmetry-breaking corrections $\delta_c$
for the $0^+\rightarrow0^+$ superallowed transitions. In the RHF+RPA
framework the density-dependent effective interactions PKO1, PKO2,
and PKO3 are employed, while in the RH+RPA framework the
density-dependent effective interactions DD-ME1 and DD-ME2 as well
as the nonlinear effective interactions NL3 and TM1 are used.

It is found that the proper treatments of the Coulomb field is very
important to extract the isospin symmetry-breaking corrections
$\delta_c$. By switching off the exchange contributions of the
Coulomb field, $E_x$ and $\delta_c$ in RHF+RPA calculations recover
the results in RH+RPA calculations. In other words, although the
meson exchange terms can be somehow effectively included by
adjusting the parameters in the direct terms, this has not been done
for the Coulomb part in the usual RH approximation.

With the isospin symmetry-breaking corrections $\delta_c$ calculated
by relativistic RPA approaches, the nucleus-independent
$\mathcal{F}t$ values are obtained in combination with the
experimental $ft$ values in the most recent survey and the improved
radiative corrections. It is found that the constancy of the
$\mathcal{F}t$ values is satisfied for all self-consistent
relativistic RPA calculations here.  It is also found that the
$\mathcal{F}t$ values of RHF+RPA are about 2~s larger than those of
RH+RPA, which are larger than the difference due to the different
effective interactions in either RHF or RH approximations.

The values of $|V_{ud}|$ thus obtained well agree with those
obtained in neutron decay, pion $\beta$ decay, and nuclear mirror
transitions. However, the sum of squared top-row elements
considerably deviates from the unitarity condition, which is in
contradiction with the conclusion in shell model calculations (H\&T)
\cite{0812.1202v1}.

For the further studies, more intensive investigations including the
proper neutron-proton mass difference, isoscalar and isovector
pairing, and deformation should be done.

\section*{Acknowledgments}

This work is partly supported by Major State 973 Program
2007CB815000, the NSFC under Grant Nos. 10435010, 10775004, and
10221003. One of the authors (H.L.) is grateful to the French
Embassy in Beijing for the financial support for his stay in France.

\bibliography{ref}

\end{document}